\documentclass[12pt,preprint]{aastex}

\begin{document}

\newcommand{\vdag}{(v)^\dagger}
\newcommand{\myemail}{aprestwich@cfa.harvard.edu}
\newcommand{\chandra}{{\it Chandra}}
\newcommand{\swift}{{\it SWIFT}}
\newcommand{\mac}{[MAC92] 17-A}
\newcommand{\xone}{IC 10 X-1}
\newcommand{\heii}{HeII 4686\AA}
\newcommand{\xmm}{{\it XMM-Newton}}
\newcommand{\kms}{km s$^{-1}$}
\newcommand{\msun}{M$_{\odot}$}


\shorttitle{The Orbital Period of IC 10 X-1 }
\shortauthors{Prestwich et al.}


\title{The Orbital Period of the Wolf-Rayet Binary \xone; Dynamic
Evidence that the Compact Object is a Black Hole}

\author{A .H. Prestwich \& R. Kilgard}
\affil{Harvard-Smithsonian Center for Astrophysics, Cambridge, MA
02138}
\author{P.A  Crowther}
\affil{Department of Physics and Astronomy, University of Sheffield, UK}
\author{S. Carpano \& A.M.T. Pollock}
\affil{\xmm\ Science Operations Center, ESAC, 28080 Madrid,
Spain}
\author{A. Zezas \&  S.H. Saar}
\affil{Harvard-Smithsonian Center for Astrophysics, Cambridge, MA
02138}
\author{T.P. Roberts \& M. J. Ward}
\affil{Dept of Physics, Durham University, South Road, Durham DH1 3LE,
UK}

\begin{abstract}
\xone\ is a bright (L$_x$=10$^{38}$ ergs$^{-1}$) variable X-ray
source in the local group starburst galaxy IC 10.  The most plausible
optical counterpart is a luminous Wolf-Rayet star, making \xone\ a rare
example of a Wolf-Rayet X-ray binary.  In this paper, we report on the
detection of an X-ray orbital period for \xone\ of 34.4 hours.  This result,
combined with a re-examination of  optical spectra, allow us
to determine a mass function for the system  $f(M)=7.8$\msun\ and a
probable 
mass for the compact object of $24-33$\msun.  If this analysis is
correct, the compact object is the most massive known stellar black
black hole.  We further show that the
observed period is inconsistent with Roche lobe overflow, suggesting
that the binary is detached and the black hole is accreting the wind
of the Wolf-Rayet star.  The observed mass loss rate of \mac\ is
sufficient to power the X-ray luminosity of \xone.

\end{abstract}

\section{Introduction}

Models of the evolution of High Mass X-ray Binaries (HMXB) predict the
existence of helium star+compact object pairs
\citep{vandenH73}.   Such systems should form at the very end of the
X-ray binary evolution, after the secondary (donor) star has been
stripped of it's hydrogen either through Roche lobe overflow or mass
loss via a strong wind.  In either case, a luminous helium star with
a compact companion is formed \citep{E&Y98}.  These systems are expected to be
rare.  \cite{E&Y98} predict $\sim$100 helium star+black hole
pairs in the Galaxy.  However, only a small number of these are
expected to form accretion disks and hence be visible as X-ray sources.  Identification of such rare systems is important,  because they have
the potential to put strong constraints on the evolution of massive binary
pairs.  There are currently only three candidates:  Cyg X-3,  NGC 300
X-1 
\citep{carpano07a} and \xone.   

\xone\ is a bright (L$_x$=10$^{38}$ ergs$^{-1}$) variable X-ray
source in the local group metal-poor starburst  galaxy IC 10
\citep{brandt97,B&B04}.  It is surrounded by a shell of non thermal
radio emission \citep{yang93} and X-ray emission
\citep{wang05,brandt97} which may be associated with the supernova
which produced the compact object in \xone.  There are 4 possible
optical counterparts  to the X-ray source,
with the most plausible being a bright Wolf Rayet star \mac\ 
\citep{Crow03}.  Spectroscopic observations of \mac\ reveal prominent
He II line emission, suggesting an identification as a WNE star
\citep{C&C04}.  In this
paper we report the discovery of an X-ray orbital period in \xone\ using data
from \swift\ and \chandra\ \citep{prestwich_atel06}.  In
 Section~\ref{obs} we describe the \chandra\ and \swift\
observations, and in  Section~\ref{discuss} we discuss the constraints
the period puts on the accretion mechanisms and mass of the compact
companion.

\section{Obervations and Data Analysis}
\label{obs}
\begin{figure}
 \epsscale{0.5}
 \plotone{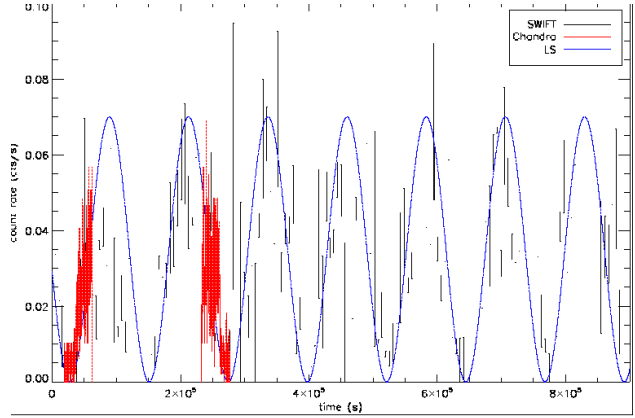}
\caption{\swift\  lightcurve of \xone\ in 100 second time bins.  Overplotted in red is the \chandra\ lightcurve in 1000 second bins.  The \chandra\ count rate is reduced by a factor of 5 to approximate the lower effective area of \swift.  The \chandra\ data is also shifted forward in time by $13\times\ p$ where p is the Lomb-Scargle period.  Times are relative to the start of \swift\ observations, 2006 21 November at 05:11:05 UTC.  The blue curve is the period derived by the Lomb-Scargle periodogram.}
\label{fig:lc}
\end{figure}

\xone\ was observed with \chandra\ on 2006 2 November and 2006 4
November for approximately 45 ks per observation (ObsIDs 7802 and
8458).  Data were processed using standard \textit{CIAO} analysis
software, version 3.4.  After extraction of a lightcurve, we observed
a large (factor of 7) flux increase in the first Chandra observation followed
by a similar flux decrease in the second observation (see
Figure~\ref{fig:lc}).  The large flux change and sharp profile of the
flux modulations suggested we were observing eclipses in \xone.
We immediately applied for, and were generously awarded, a SWIFT
target-of-opportunity (TOO) observation of 100 ks.  \xone\ was observed with \swift\ beginning 2006 21 November for
approximately 700 seconds per 90 minute \swift\ orbit for a total of
97 ks spanning 246 hours.  Data were processed by the \swift\  pipeline and screened using standard procedures outlined in the XRT Data Reduction Guide.

\subsection{Timing Analysis} 

\begin{figure}
 \epsscale{0.5}
 \plotone{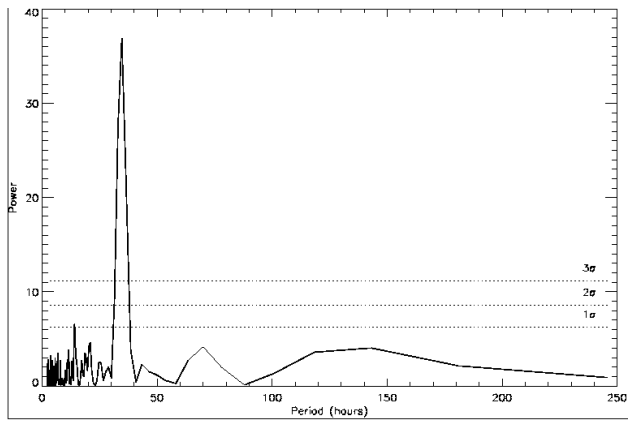}
\caption{The power spectrum of \xone.  The horizontal dotted lines
correspond to the false-alarm probability corresponding to 1, 2, and
3-sigma significance.}
\label{fig:PG}
\end{figure}

In order to search for periodic signals in the \swift\ observations of
\xone, we merge the \swift\ event files using the \textit{CIAO} tool
\textit{dmmerge}.  We extract the events in a 60\arcsec\ radius around
the \chandra\ source position; this is reasonable due to both the
extent of the \swift\ point-spread function and the sparsity of nearby
sources as observed in the \chandra\  observations.  The A Lomb-Scargle (LS)
periodogram of \xone\ is shown in Figure~\ref{fig:PG}.  The
periodogram shows no evidence for increased noise at low frequencies
(``red noise''), indicating that the LS method can be used to search
for peaks in the power spectrum.  The
full data range was searched from the Nyquist frequency (twice the
2.5073 s XRT readout rate) to half the full observation duration of
123 hours.  The normalization used in the LS method is the
total variance of the data \citep{horne86}.  The LS periodogram determines a period of 34.40.  Following the method of
\cite{horne86} we determine a period
uncertainty of $\pm$ 0.83 hours. 

 Two sets of simulations to were
carried out to ascertain the false alarm probability (FAP) of the signal. The first
was a Monte Carlo simulation, in which fake spectra were generated
with  Gaussian noise set to the standard deviation of the actual data.   The second was a bootstrap method,
which creates a fake periodogram by randomly rearranging real data
values.  Observation times were taken from the real data, and
$1\times10^5$ fake spectra were generated  for each method.   The number of simulated spectra  with peak amplitudes greater
than the the observed peak amplitude (i.e. the FAP) was
0 in both Monte Carlo and Bootstrap simulations.  Thus the FAP $< 1\times10^{-5}$,
and the significance $> 4.5 \sigma$

 \xone\ was observerd in 2003 by both \chandra\ and \xmm.  The 30 ks \chandra\ observation of \xone\ showed the source to be
variable, but did not show any evidence for an eclipse \citep{B&B04}.
A sharp increase in the X-ray flux of \xone\
during the 45 ks \xmm\ observation is probably due to an eclipse egress
\citep{wang05}.  The data in these observations are consistent with
the period derived in this paper.

\section{Discussion}
\label{discuss}

 The most plausible explanation for the
regular flux modulations is an
orbital eclipse of the X-ray emitting object by the donor
star.  It is unlikely the observed period is super-orbital (e.g. due to a precessing
warped accretion disk)  because it is shorter than
most super-orbital periods (most are 10-100 days,
e.g. \cite{clarkson04}).  While we cannot rule out an alternative
explanation without a fully sampled  optical radial velocity curve, we
assume the modulation is orbital for the rest of the paper.

\subsection{Constraints on the Accretion Mechanism}

The period can be used to constrain the accretion mechanism, in
particular to determine whether the donor fills its Roche lobe, or
whether accretion occurs via a wind.   Here
we test the hypothesis that the accretion is via Roche lobe overflow
by determing whether the observed period is consistent with
theoretical values of the period for compact object masses in the
range 1-100\msun.

The orbital period, P, is related to the mass of the primary ($M_1$),
the mass of the donor ($M_2$)  and
the separation between them ($a$) via Kepler's Third Law:

\begin{equation}
\label{kep}
P=\frac{2\pi a^{3/2}}{[G(M_1+M_2)]^{1/2}}
\end{equation}

Here we assume that the inclination of the system is close to 90 $\deg$, as
implied by the existence of an eclipse.  
If we assume that the donor fills its Roche lobe,  we can make the
simplifying assumption that the Roche lobe radius ($R_{cr}$) is equal
to the donor radius ($R_2$).  The separation can then be expressed as
\citep{egg83}:

\begin{equation}
a=R_{2} \frac{0.6q^{2/3}+ln(1+q^{1/3})}{0.49q^{2/3}}
\label{roche}
\end{equation}

where $q=M_2/M_1$.  If the mass and radius of the donor star are
known, equations \ref{kep} and \ref{roche} can be used to determine
values of the period for plausible masses of the compact object. 

 The
mass of \mac\ derived from spectroscopic data is $\sim35$ \msun\
\citep{C&C04}.  The uncertainty on this value is large.  It is
possible (but unlikely) that \mac\ might have a mass as low as 17
\msun.  The radius of \mac\ is derived using standard Mass-Radius values for WR stars taken from
\cite{lang89}. Assuming that \mac\ is indeed the donor star, we find
that for M$_1$=1-100\msun\ the period derived using equations
\ref{kep} and \ref{roche} is  2-3.5 hours.   These conclusions still
apply if the mass of \mac\ is closer to 17 \msun than 35 \msun.  Clearly, the observed
period is inconsistent with Roche lobe overflow and implies that
accretion is via a wind as seen in many other high mass X-ray
binaries.  

\subsection{The Mass of the Compact Object}

The standard method to measure the mass of stars in a binary system
is to determine the mass function (e.g. \cite{mcl06}):
\begin{equation}
f(M) \equiv \frac{P K_{2}^{3}}{2\pi G}=\frac{M_1\sin^3 i}{(1+q)^2}
\label{mf}
\end{equation}

where $K_2$ is the half amplitude of the velocity curve of the
secondary.  The mass function is the minimum mass of the compact
primary.  In the case of the \xone\ system, we have reasonable
estimates for the the orbital period and the 
inclination but require  a value of $K_2$ to secure the mass
function.    The centroid of the \heii\ line is commonly used to
determine the radial velocity of WR+OB  binaries because it is formed
in the inner wind, close to the star.   In the light of the discovery of the orbital period in 
\xone, we re-examined the optical spectra presented by \cite{C&C04} to
search for shifts in the centroid of the  \heii\ line which might
be due to the orbital motion.  We note that the spectrum of \mac\ is
characteristic of a WN star, and does not show evidence for features
seen in accretion driven outflows (e.g. HI, HeI, FeII).  We are
therefore reasonably confident that the \heii\ line in \mac\ is associated with the
donor star and has minimal contribution from any accretion disk.

\begin{figure}
\epsscale{0.5}
 \plotone{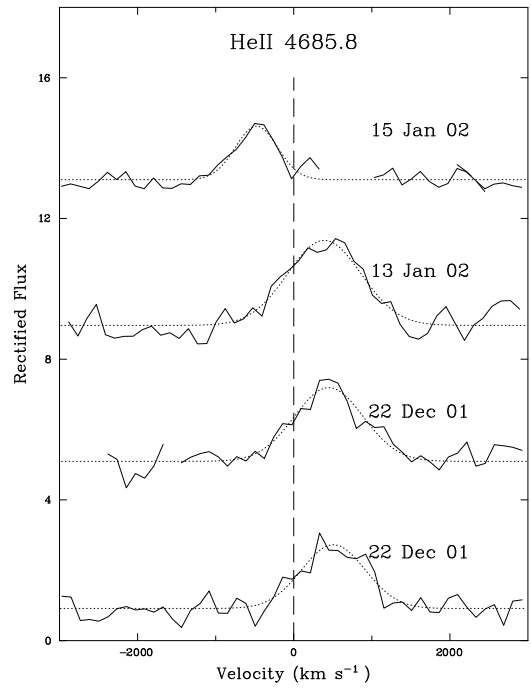}
\caption{The HeII 4686 \AA line in \mac.  Four individual spectra are
shown. Zero velocity is shown as the dashed line.  There is clear evidence that the
line centroid has shifted.}
\label{fig:opt}
\end{figure}

 Individual Gemini GMOS spectra of \mac\ were re-extracted
relative to the IC~10 nebular [OIII]  5006.8 \AA\ line. 
  Figure~\ref{fig:opt} shows the individual  spectra together with
Gaussian fits to the \heii\ line.  Two one-hour exposures were taken sequentially on the
night of December 22 2001, and two further spectra obtained on the nights of
January 13 2002 and January 15 2002.  As might expected for a binary
system with a period of 34.4 hours, there is no evidence for a shift
in the velocity centroid of the stellar  \heii\ line within IC 10 for the two
exposures taken on December 22 2001.  However, the line centroid on 15
Jan 2002 is definitely shifted substantially relative to the other
datasets.  The FWHM of the fitted gaussian for the 15 Jan 2002 dataset is narrower
than that determined for the other epochs (10\AA\ vs 15-17\AA).  This
may be due to a CCD defect which contaminates the  red component of
the 15 Jan 2002 line.  Our best estimate for the centroid of the \heii\
line on 15 Jan 2002 is $\sim -300$ \kms, compared to $\sim
+450$ \kms\ for the other epochs.

We make the assumption that the
shift in the \heii\ line centroid  of approximately $750$ \kms\ is
due to the radial velocity of \mac.  We further assume that
$750$ \kms\ corresponds to the maximum possible  velocity
displacement.  We find that $K_2=375$ \kms\ and the mass function
$f(M)=7.8$ \msun.  This value of the mass function implies that the compact
object is a black hole.  

We can further constrain the mass of the
compact object by using equation \ref{mf} and assuming a range of 
reasonable values for the mass of the donor star and inclination of
the system, as shown in Table~\ref{tab:inc}.  The mass of \mac\ derived by \cite{C&C04} from
spectroscopic data is  $\sim35$ \msun.  There is considerable
uncertainty in this estimate, and so we use a conservative lower
limit of  $17$ \msun.  The existance of deep eclipses suggests that the
inclination of the system is close to 90 $\deg$, however we cannot rule
out the possibility that the inclination is considerably less and use
45 $\deg$ as a lower limit.  Table~\ref{tab:inc} demonstrates that the
minimum mass of the black hole to satisfy the constraints of
equation~\ref{mf} is $\sim23$ \msun.  Assuming the inclination of the
system is close to 90 $\deg$, the mass of  the black
hole is most plausibly in the range $23-34$ \msun.  However, if the
inclination is much lower than expected (close to 45 $\deg$) the black hole mass can be as
high as $\sim57$ \msun.  

If the mass of the the compact object in \xone\ is indeed $>$23 
\msun, it is the most massive known stellar black hole with a
dynamically a determined mass (the record is curently held by GRS
1915+105 with a mass 10-18 \msun \citep{r&mcc06}.)  Most models
predict that the remnants of a supernova explosion have masses below
20 \msun \citep{Fry&K01}.  It is possible that the black hole in \xone\ has increased
it's mass by accreting material from \mac\ \citep{Pod03}  The low metallicity of IC
10 may also be conducive to the formation of higher mass black holes \citep{Fry&K01}. Detailed
theoretical models need to be developed to understand the evolutionary
history of \xone.

\subsection{Is the Model of \xone\ Self Consistent?}
\label{consist}

The results from this and previous studies of the \xone\ system suggest
that it is a detached binary composed of a massive
($\sim35$ \msun) Wolf-Rayet star with a  23-34 \msun\ black hole
companion.  X-rays are produced as  material from the wind of \mac\ is
accreted via an accretion disk onto the black hole.  It is important to test the self-consistency of this model
and show that such a system is indeed capable of forming an accretion
disk and producing X-rays.

A black hole in a detached binary system can form an accretion disk 
(and hence be a strong X-ray source) if the following condition is
met e.g.\citep{E&Y98}:

\begin{equation}
P < 4.8(M_{BH}/M_{\odot})v_{1000}^{-4}\delta^2
\end{equation}
where $M_{BH}$ is the mass of the black hole, $v$ the velocity of the
wind impacting the compact object in 
units of $1000$ \kms, P is the period in hours, 
 and $\delta$ a dimensionless parameter of order unity.  If the mass
of the black hole is $24-34$ \msun\ we find that this condition is
satisfied if $v \leq 1400-1500$ \kms.  The terminal wind velocity of
\mac\ was estimated by \cite{C&C04} to be  $v_{\infty}= 1750$ \kms.  The velocity of the wind impacting the black hole will be
less than the terminal velocity; the upper limit of $1400-1500$ \kms\
is consistent with the wind structure of a Wolf-Rayet star
\citep{carp07}.

The energy released by material falling onto a compact object via an
accretion disk is given by $L=\eta \dot{M} c^2$, \citep{S&S73},
where $\dot{M}$ is the accretion rate and $\eta$ the efficiency.  If
we assume  $\eta=0.1$, the mass accretion rate required to sustain an
X-ray luminosity of  L$_x\sim 2\times10^{38}$ ergs$^{-1}$ is
$\sim 3.5 \times 10^{-8}$  \msun yr$^{-1}$.  The
mass loss rate of \mac\ derived by \cite{C&C04} is $\sim 10^{-5}$ \msun
yr$^{-1}$.  We therefore conclude that direct wind fed accretion is
consistent with the observed X-ray luminosity.
 
\subsection{\xone\ and the Star Formation History of IC 10}

Both the mass of the donor and the mass of the compact object in
  \xone\  indicate that
  this system is less than 10~Myr old.  This is consistent
  with the suggestion that there have been pockets of
  star formation $\sim3-4$~Myr ago  which may still be on-going
  \citep{hunter01}.  There are several dozen WR stars in IC 10, also
  indicating  recent star formation 
  \citep{M&H02,Crow03}.  The fact that the donor is a
  He-core WR star indicates that its initial mass was much higher
  (probably   in
excess of 60\msun\ \cite{M&M05}) in which case  the
system could have formed in one of the recent star formation events. 
 The presence of a shell of ionized gas around the system, probably
related to supernova activity, indicates that  there is still on-going
star-formation in that region.   The formation of  the very massive progenitors of this binary requires a very
flat IMF.  \cite{hunter01}, finds that such a flat IMF is consistent
with a star-formation scenario of widespread activity $\sim40$~Myr ago
and more recent ($\sim3-4$~Myr) localized events. 

\section{Summary and Conclusions}
We have used \swift\ and \chandra\ to determine  the orbital period of
the bright Wolf-Rayet X-ray binary \xone\ to be 34.4 hours.  Using this
value of the period, and the mass of the donor star derived by
\cite{C&C04} from optical spectroscopy, we  use Kepler's laws to show
that the donor star almost certainly does not fill its Roche lobe.
We cannot derive the mass function of the binary system or mass of the
compact companion from the period alone.  However, re-examination of
published optical data reveals a shift in the centroid of the
\heii\ line  most likely due to orbital motion of the donor
star.  If this is the case, the mass function  $f(M)\sim7.8$ \msun\ and
the compact object is a black hole.  Assuming that the mass of the
donor star is $17-35M$ \msun, we calculate that the black hole mass is
$\sim24-34$ \msun.  If confirmed, this makes \xone\
the most massive black hole binary known.  We
show that it is possible for an accretion disk to form and that the
observed mass loss rate can power the X-ray luminosity.

\section{Acknowledgments} 
We are grateful to the Swift science team, in particular Neil Gehrels
and David Burrows for approving our Swift TOO.  We also thank Swift
team members Kim Page and David Morris for assistance with observation
planning and analysis.  We are very  grateful to Miriam Krauss for
help with the timing analysis, and to
Frank Primini and Jeff McClintock for helpful discussions.
  
Support for this work was provided by the National Aeronautics and
Space Administration through Chandra Award Number  GO6-7080X issued by
the Chandra X-ray Observatory Center, which is operated by the
Smithsonian Astrophysical Observatory for and on behalf of the
National Aeronautics Space Administration under contract
NAS8-03060. This work was also supported by NASA contract NAS 8-39073.

This paper
uses data from the Gemini Observatory, which is opperated by the Association of Universities for Research in Astronomy, Inc., under a cooperative agreement with the NSF on behalf of the Gemini partnership: the National Science Foundation (United States), the Particle Physics \& Research Council (UK), the National Reseach Council (Canada), CONICYT (Chile), the Australian Research Council (Australia), CNPq (Brazil), and CONICET (Argentina).

\clearpage

\begin{deluxetable}{ccccc} 
\tablecolumns{5} 
\tablewidth{0pc} 
\tablecaption{Black Hole Mass (in \msun) as a Function of Inclination and  Donor Mass} 
\tablehead{ 
 \colhead{} & \colhead{} & \multicolumn{3}{c}{Donor Mass (\msun)} \\
\colhead{} & \colhead{}  & \colhead{17} & \colhead{25}   & \colhead{35}\\
\colhead{Inclination ($\deg$)} & \colhead{} & \colhead{} & \colhead{} &\colhead{} }
\startdata 

 90&   & 23 & 28 & 34\\
60 &  & 29 & 35 & 41\\
45 &  & 43 & 49 & 57 \\
\enddata 
\label{tab:inc}
\end{deluxetable} 

\clearpage

\end{document}